# Nonstationary behavior of the upconversion processes of $Er^{3+}$:$Yb^{3+}$ ions pair doping the $Y_2O_3$ ceramic illuminated with 976 nm laser light

**Liviu Dudaș**

Faculty of Chemical Engineering and Biotechnologies, National University of Science and Technology Politehnica Bucharest, 1-7 Gheorghe Polizu Street, 011061 Bucharest, Romania; liviu_dudas@yahoo.com (L.D.);

**Abstract**: The studies of the upconversion response of RE ions in various hosts usually depict a static picture, whereas, at constant illumination, the upconverted emission is also constant. We discovered here that, at least in the case of $Y_2O_3$ doped with $Er^{3+}$:$Yb^{3+}$, this is not the case. At constant illuminated power, the upconversion emission is not only fluctuating, but it varies in a manner which seems to be a superposition of intensity waves with different frequencies. We present here the results of the measurements we've performed on $Y_2O_3$ with 1%$Er^{3+}$-2%$Yb^{3+}$. The origin of this effect remain to be established but it shows that the energy transfers during the upconversion are an interplay of absorption-saturation-emission-relaxation processes. (This data was presented at Nanobiomat 2023 conference).

## 1. Introduction

The pellets of $Y_2O_3$ (YO) 1-2 described in [1] were illuminated with IR laser light at 976 nm and the upconversion to visible range (500 - 700 nm) spectra were measured. The experimental setup is shown in Figure 1.

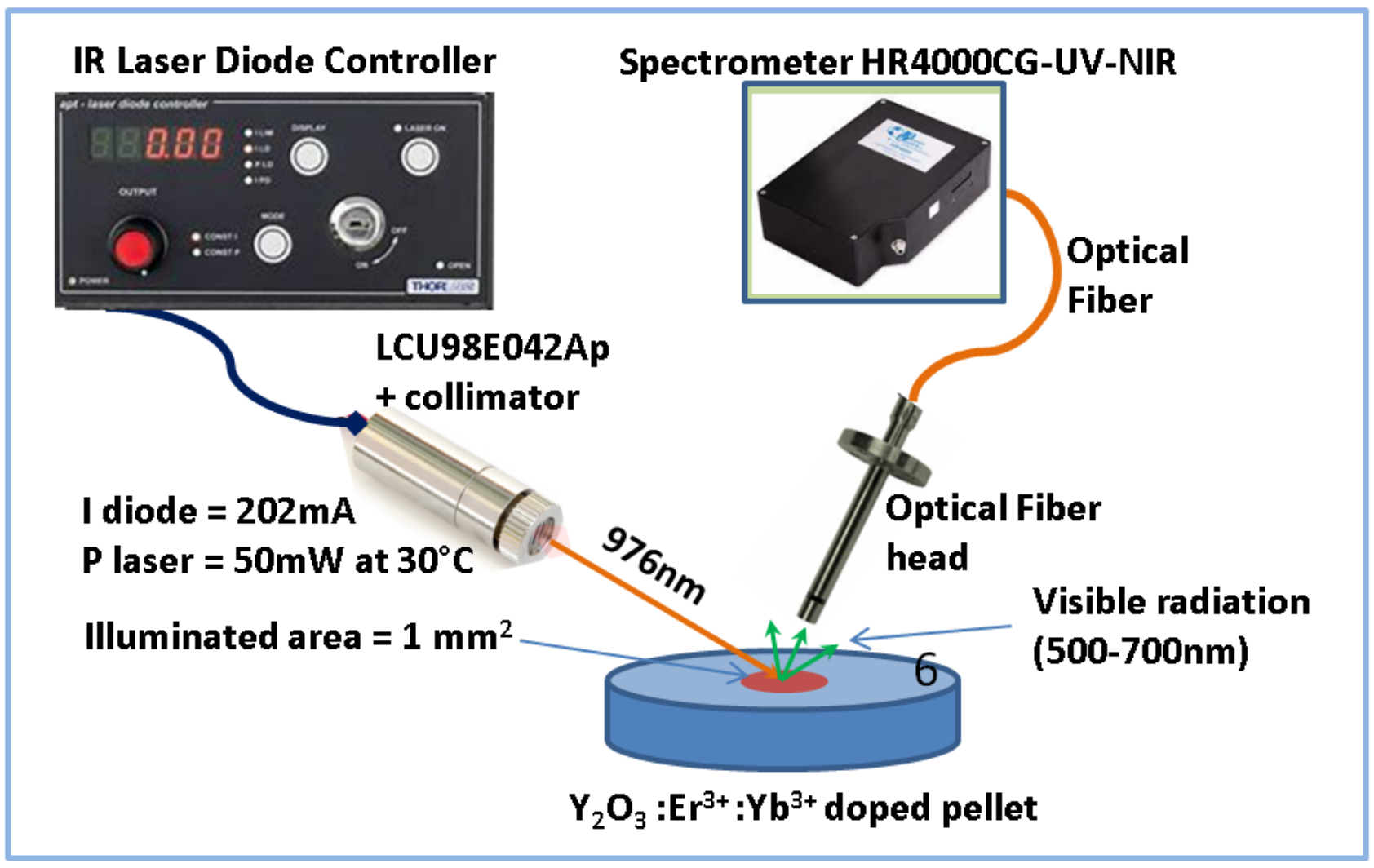

**Figure 1**. The measurement setup for the ceramic pellet (from [1]). The illumination power was maintained constant at 50 mW and the pellet's upconverted emission was measured during cooling to room temperature, warming to room temperature or being at constant room temperature.

The spectra measurements were done over a duration of minutes, maintaining constant the illumination power at 50 mW on a surface of about 1 mm² (power density 5 W/cm²). The temperature of the pellets had three regimens, constant at room temperature (25 °C), warming from 0 °C to room temperature and cooling from 70 °C to room temperature.

Only the starting temperatures of the pellet were measured, during the process, the pellet temperature was neither monitored or controlled. This was because this experiment is only an evaluative one only seeking to asess the phenomenon.

The resulting spectra were split into four parts as is shown in Figure 2, each part corresponding to emission from the specified transition of $Er^{3+}$. While GRNA and GRNB are from distinct transi-tions, REDA and REDB are decays from the crystal field splitted sublevels of $^4F_{9/2}$ level. As such, sinsce REDA and REDB emissions come from the same energetic manifold, they should behave in a similar manner.

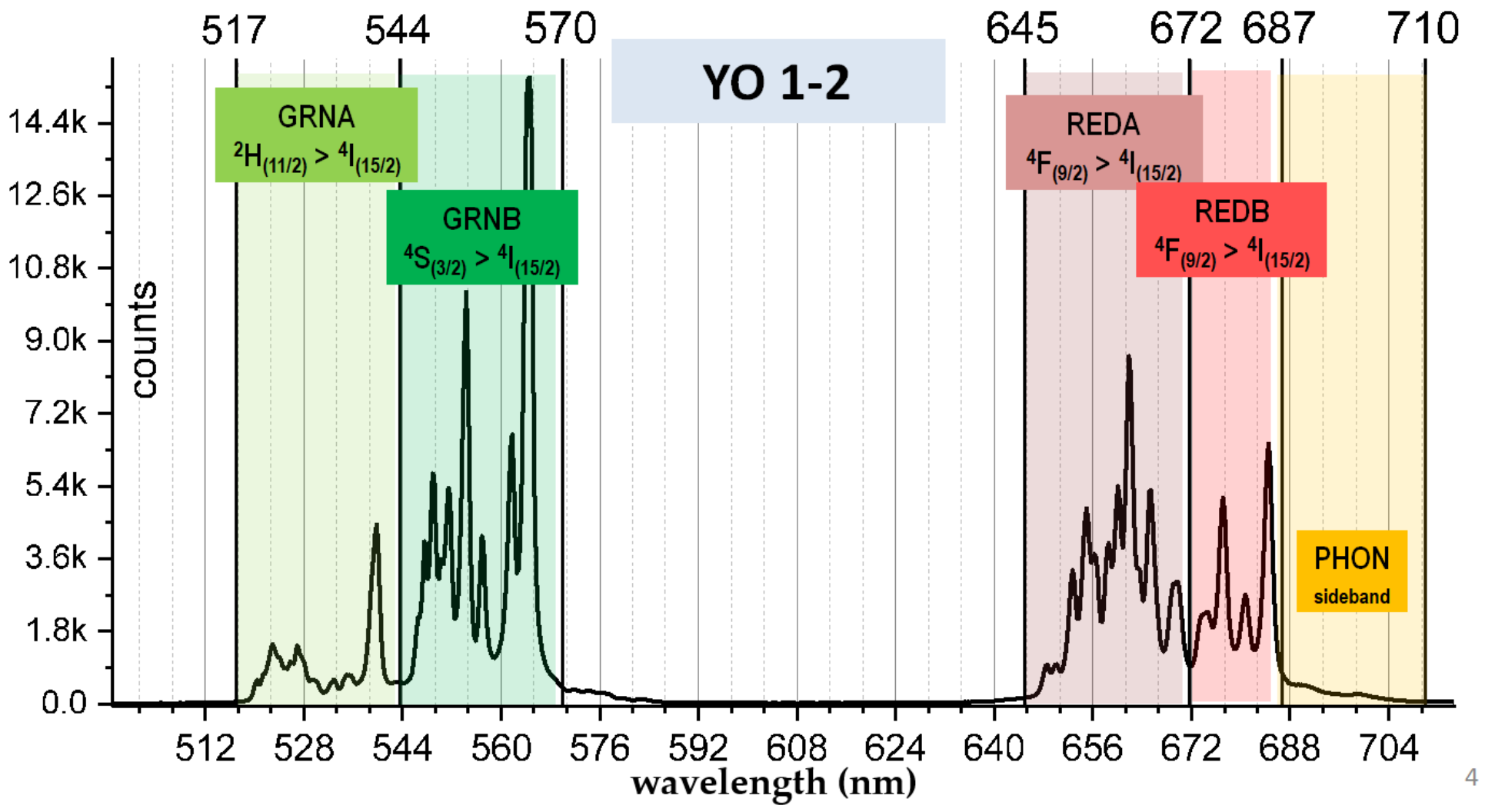

**Figure 2**. The counts for the resulted spectrum were integrated over the colored wavelength ranges depicted here, corresponding to the emissions from the respective transitions.

The reason for the spectrum splitting into regions is for better assess the each individual transition intensity and to compare their relative ratios during the time variation.

## 2. Results: monitoring the emission during cooling from 70 °C to 25 °C

Figure 3 shows how the integral emission counts for the regions depicted in Figure 2 varies during cooling starting from 70 °C. The temperature was not monitored while cooling, only the starting temperature was measured.

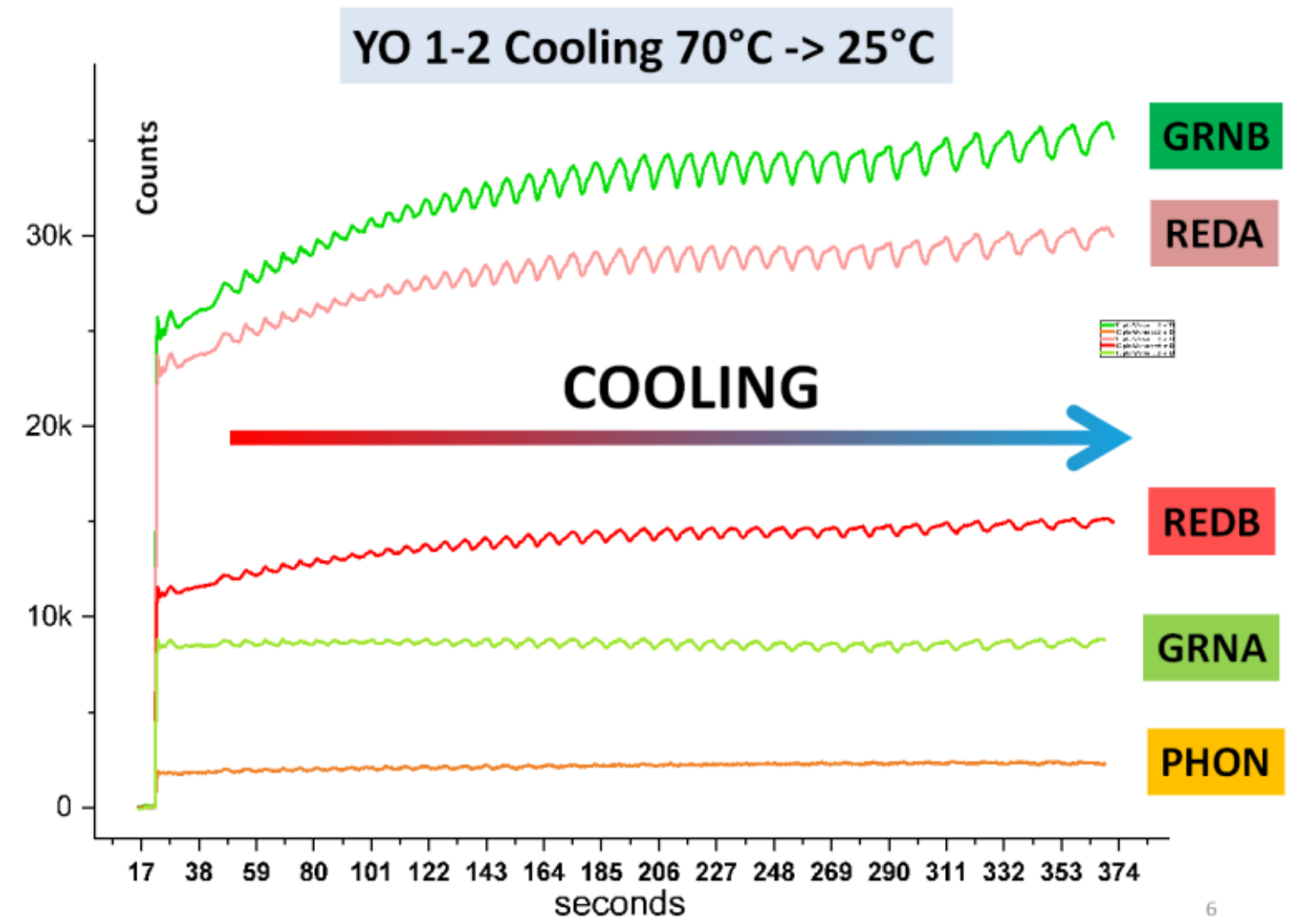

**Figure 3**. After the pellet was heated at 70 C it was quickly inserted into the measurement setup (depicted in Figure 1) and the signal monitored during the cooling to the room temperature. Heating is hindering the upconversion, that's why, at the starting moment, the intensities are low, but when the temperature of the probe is diminishing, oscillations appear. The Stokes phonon sidband emission is constant, which shows that only a relatively constant number of photons interacts with the lattice, disregarding their local density. Also, remark how the average distance GRNA ↔GRNB evolves in an exponential way.

## 3. Results: monitoring the emission during warming from 0 °C to 25 °C

Figure 4 shows how the integral emission counts from Figure 2 are varying during time. The emission is higher at the beginning, since the probe is at low temperature, which promotes the upconversion because the thermal phonons are not sufficiently dense to induce the higher levels decays for $Er^{3+}$.

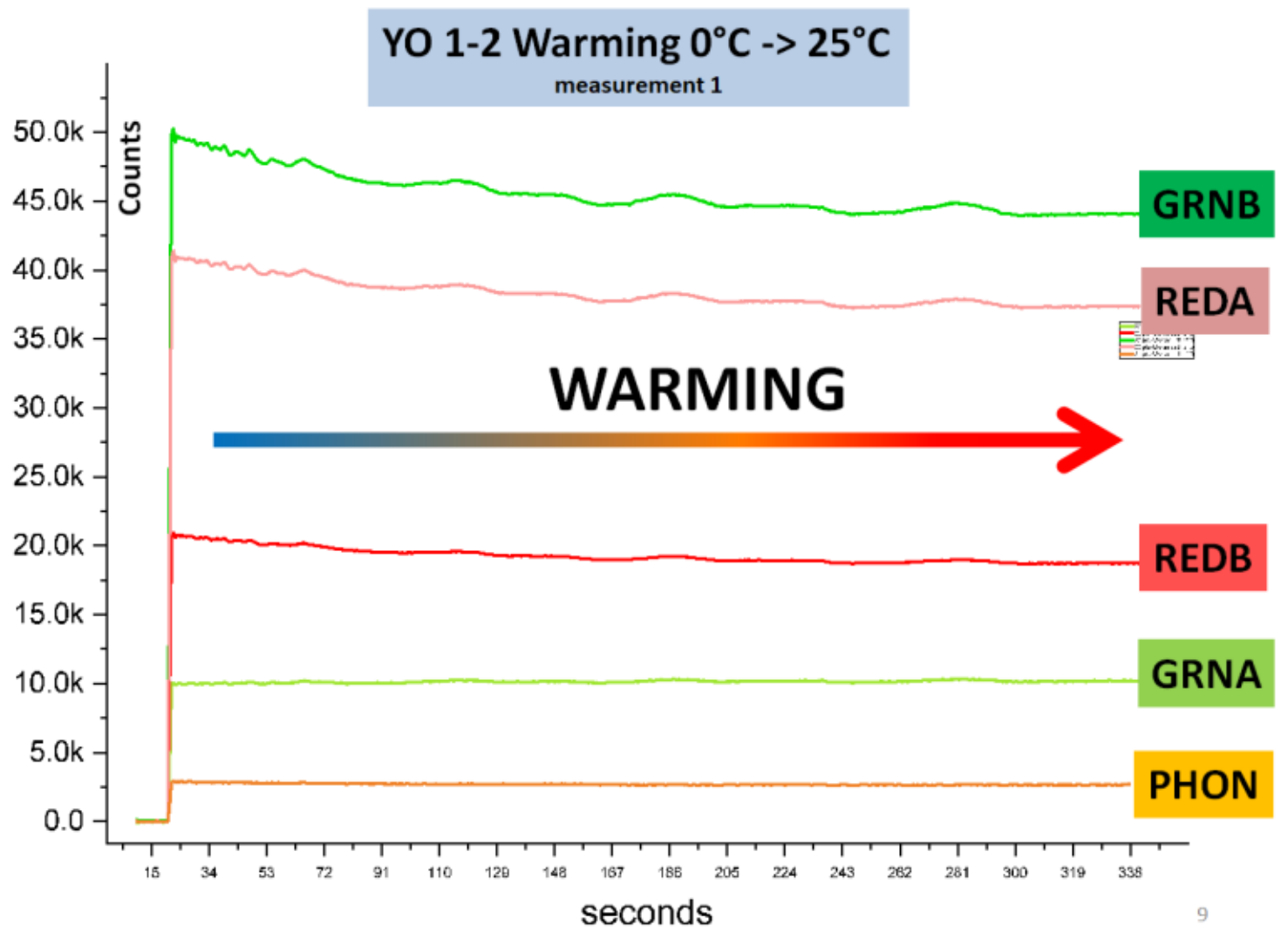

**Figure 4**. The pellet was chilled at 0 °C and then quickly inserted into the measurement system. The high initial intensities are a consequence of the lower density of thermal phonons and, as such, the life durations of the populations on the higher levels are prolonged and, by this, the decay to ground state is more intense.

## 4. Results: monitoring the emission during cooling from 70 °C to 25 °C for prolonged duration

Figure 5 shows the behavior for the emission for a longer duration than that in figure 3, the aquisition time being about 1200 sec. There are peak in the emission intensity and the relaxation is visible by decreasing the frequency of the oscillation. A FFT was performed for an interval of 300 seconds of oscillations and the result is shown in Figure 6. The highest amplitude is for 38.3 mHz and.

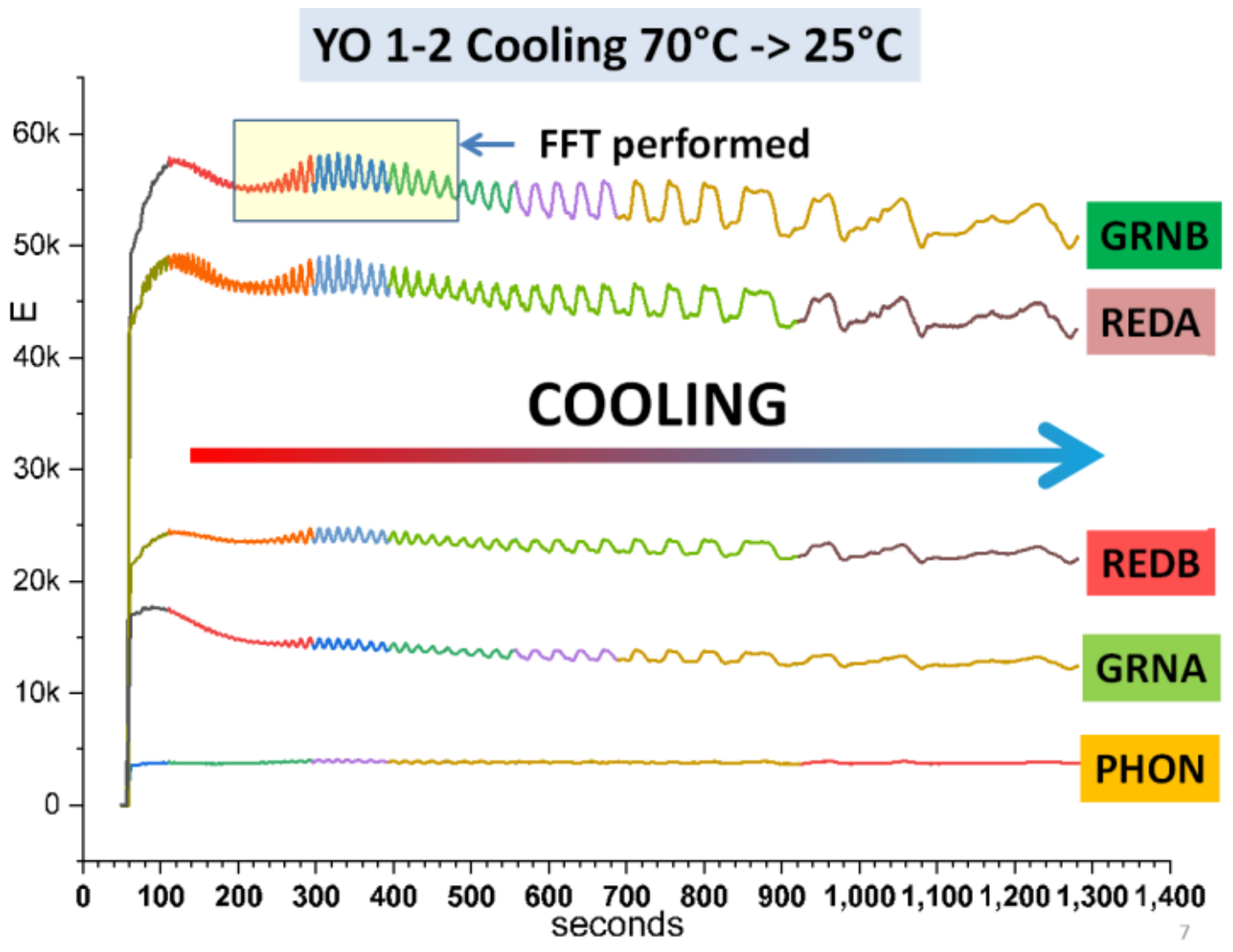

**Figure 5**. A longer cooling duration was monitored, the pellet had time to cool down to room temperature and a FFT analysis were performed for the smoothed signal in the yellow box.

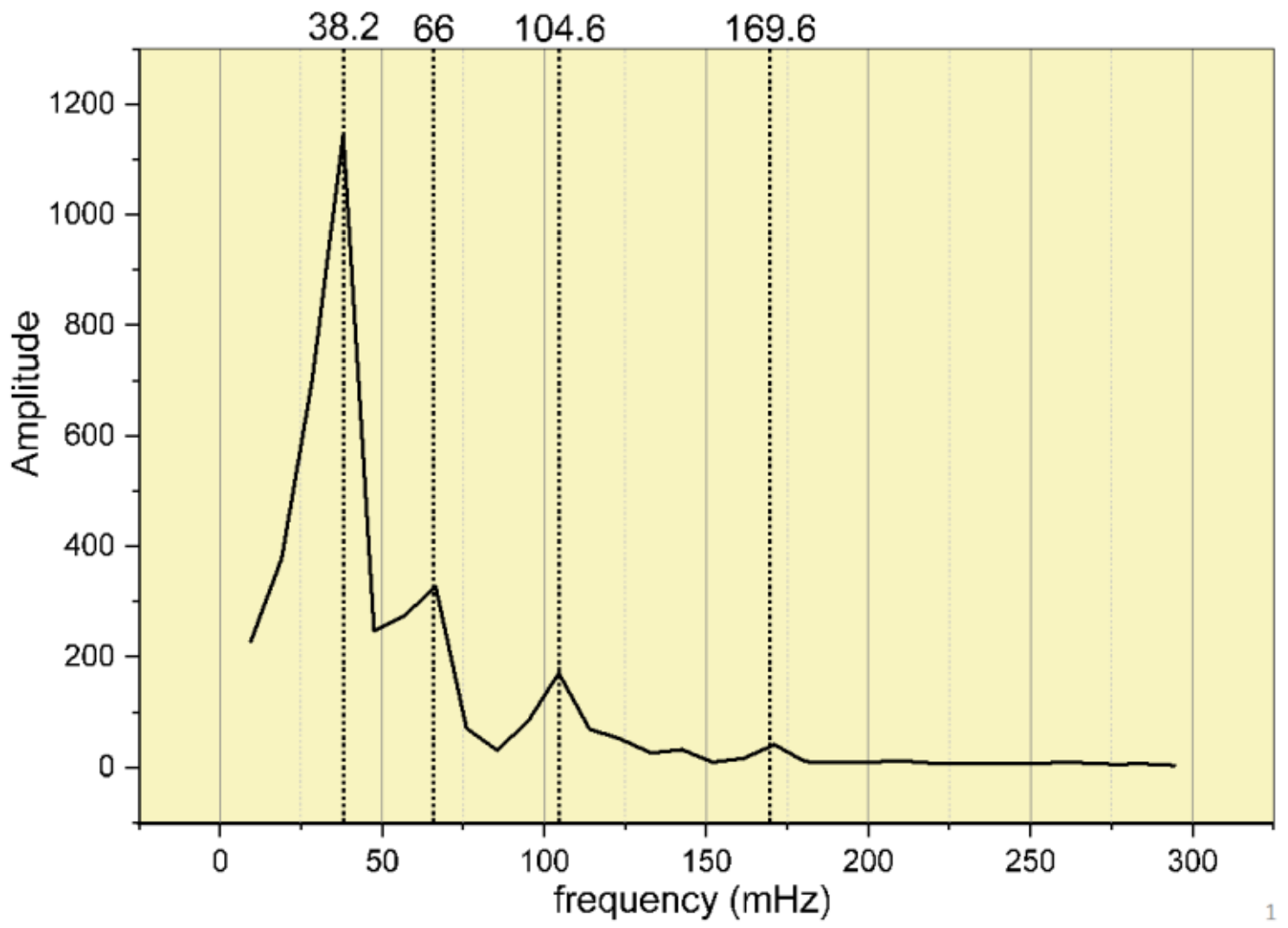

**Figure 6**. The FFT analysis of the signal portion in the yellow box in the Figure 5. The dotted vertical lines are the main spectral compontents of the oscillations. The highest amplitude is for 38.3 mHz.

## 5. Results: monitoring the integral visible emission of the probe at 30 °C, for 30 minutes

The YO 1-2 probe was also monitored for prolonged time, for about 30 minutes, at 30 °C.

This time the integral for all visible range (510 - 700 nm) was performed and the data was stitched together from chunks of 10 minutes length (red, green, blue coloring in figure 7).

In order to demonstrate the lack of spurious external influences, the anti-Stokes band of $Yb^{3+}$, situated between 900 nm and 940 nm was monitored (figure 8).

It can be seen from Figure 7, lower graph, that this integral is almost constant during the monitoring interval, which shows that the variability in the upconverted spectrum comes solely from $Er^{3+}$. One also can see, in Figure 7, the superposition of low and high frequencies of oscillation, and also the general amplitude of the variations, which is relatively high, about 20%, of the total emission..

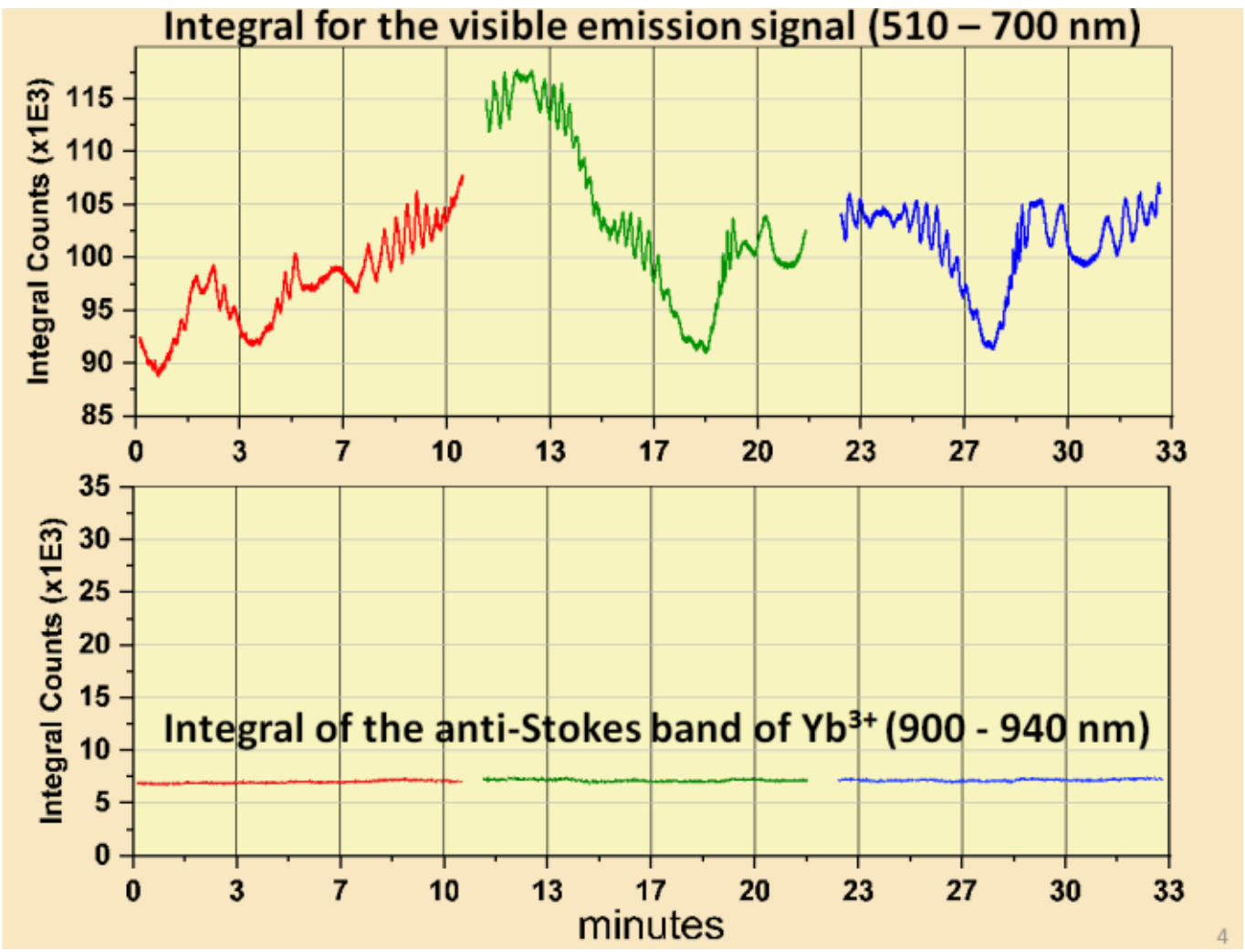

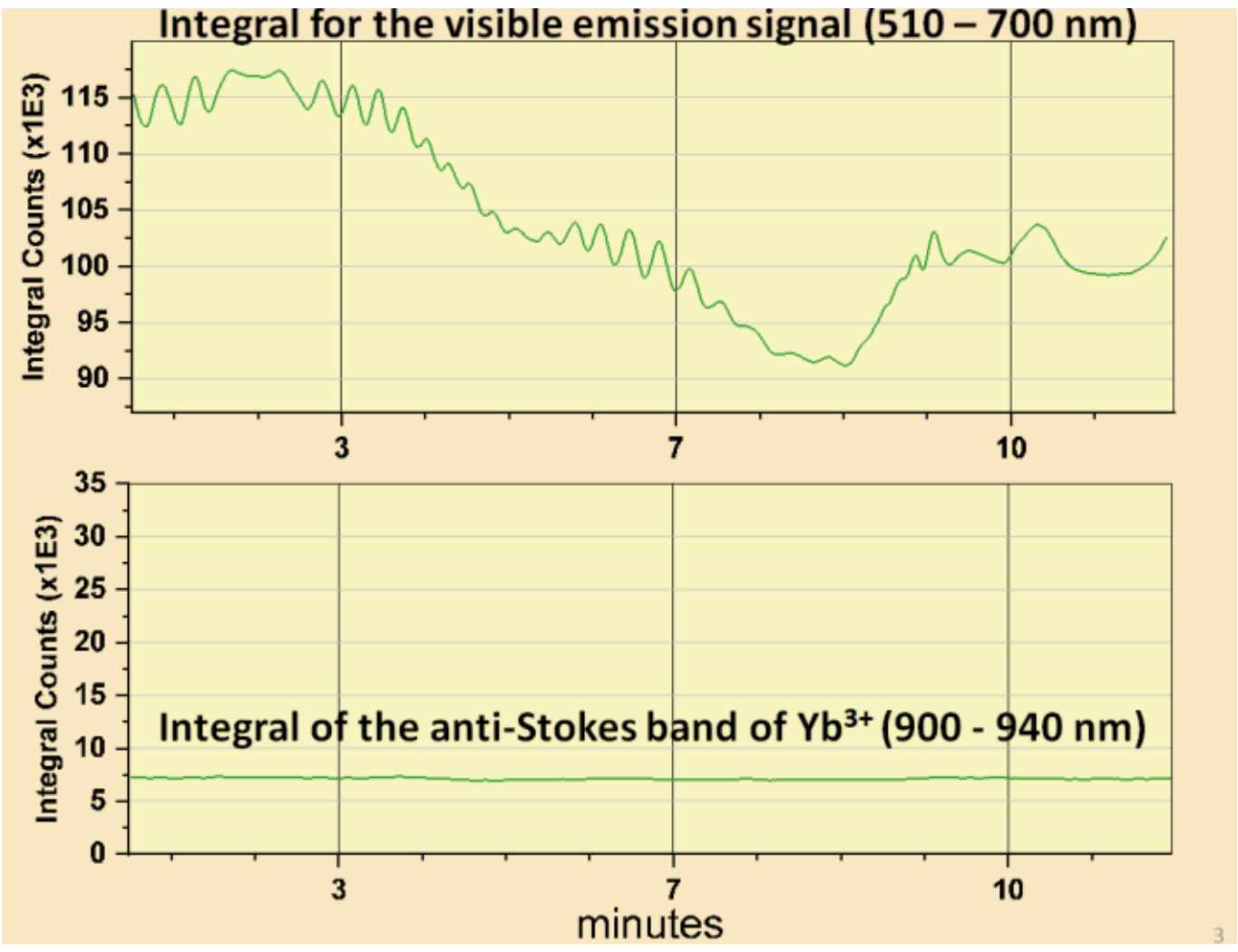

**(A)** **(B)**

**Figure 7**. (A): Integral visible emission monitored for about 30 minutes in chunks of 10 minutes stiched together. (B): The middle part of (A), duration of 10 min. Also the behavior of the integral of the $Yb^{3+}$ anti-Stokes band (between 900↔940 nm) is almost constant, which is an indication that the oscillations in the visible signal are solely due to $Er^{3+}$ ions.

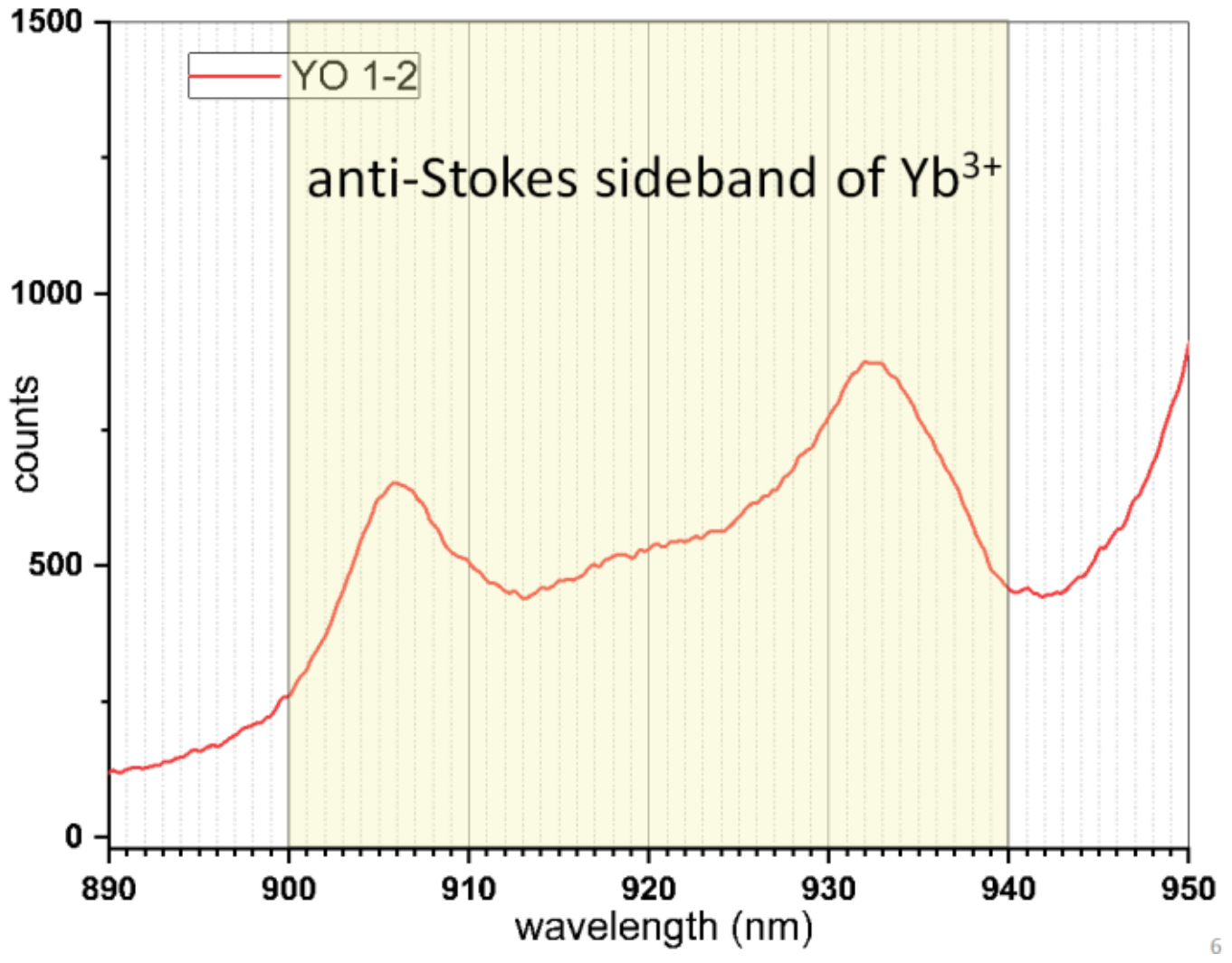

Figure 8. The anti-Stokes sideband of $Yb^{3+}$ for the emission $^4F_{5/2}\rightarrow{}^4F_{7/2}$, which is monitored in figure 7. The integral emission of this band should be independent of the oscillations of $Er^{3+}$, which is indeed the case, and its constancy shows that no external (or internal) factors were influencing the measurement, underlining that the phenomenon is solely generated by $Er^{3+}$ energy interactions.

## 6. Discussion

This behavior seems unusual, it resembles the one described in [3] but it is not the same as there, and since we didn't find any explanation in the literature, we can only speculate upon it:

- Rate equations have variable coefficients, which lead to oscillatory solutions. These coefficients could depend on the energy density of the local electric field induced by the incoming radiation and energy trapping at the nanoscale level
- Microcrystalites of the ceramic influence one another in a looping way.
- Local temperature gradients could lead to fluctuation in the ion-phonon interactions.
- Local variation of the relaxation times for $Er^{3+}$ due to variations in the saturation of the excited populations of $Yb^{3+}$ sensitizers and/or fluctuations of the power density of the local electric field [4].
- **The most appealing reason, taking into account that the activator-sensitizer interaction is somehow the same across the types of the host matrix (as explained in [1,2]), is that the basis of the effect could be a superposition of Rabi oscillations of an $Er^{3+}$ ion surrounded by $Yb^{3+}$. In this view, the activator ions, $Er^{3+}$, are trapped in cavities formed by the $Yb^{3+}$ sensitizers, and the phenomenon could be explained by the cavity QED**.

It could be the method of excitation, pin-pointing the laser light, which is very local in nature, that made this observation possible because, usually, on most of the experiments, the probe is illuminated on a wide area and, because of this light spreading on the surface, the local behaviors are averaged out, and this is the reason why no reports on this issue are to be found.

## 7. Conclusion

The spectra were split in the ranges depicted in the figure (named as shown), each range corresponding to the emission due to the transition from the excited specified level of $Er^{3+}$ to the ground state $^4I_{15/2}$.

The integral for each range, i.e., the intensity of the respective emissions, was measured with a period of 50 ms, and the data points were gathered for as long as 1200 seconds.

Initially, the intensities were expected to be constant in time; nevertheless, they displayed oscillations. To further search this behavior, the pellets were submitted to either heating to 70 °C or cooling to 0 °C prior to inserting them in the measurement setup, and the upconversion intensities were measured during the cooling (warming) to the room temperature, which was 25 °C. The exact temperature of the probes was not monitored, only the general behavior being investigated.

The oscillations are correlated with the probe's temperature, as seen in the pictures for the case of YO 1-2. The cases with the other relative $Y^{3+}$:$Er^{3+}$:$Yb^{3+}$ concentrations display the same kind of behavior.

While cooling promotes the upconversion, heating prevents it through the multiphonon non-radiative decays. It should be noted that the intensity of the transition from $^2H_{11/2}$, GRNA, is almost constant with the temperature, while only the intensity of $^4S_{3/2}$ varies, the relationship between GRNA and GRNB resembling an exponential one, as it should be, because $^2H_{11/2}$ is thermally populated from $^4S_{3/2}$. Why GRNA is almost constant is a question that further research will try to answer.

This oscillatory phenomenon can have some applications like low-frequency generators, cheap method for modulating a laser in an exterior control loop, or, if controlling its higher frequencies, burst signal generators.

EOF